\documentclass[reprint, amsmath, amssymb, aps, superscriptaddress, nolongbibliography]{revtex4-2}
\usepackage{amsfonts
}
\usepackage{amsmath,float}
\usepackage{amssymb}
\usepackage{graphicx}
\usepackage{color}
\usepackage{braket}
\usepackage[utf8]{inputenc}
\usepackage{hyperref} 
\usepackage{ytableau, comment}

\def \be {\begin{equation}}
\def \ee {\end{equation}}
\def \ba {\begin{aligned}}
\def \ea {\end{aligned}}
\def \bea {\begin{eqnarray}}
\def \eea {\end{eqnarray}}

\begin{document}

\title{Systematic Construction of Kramers-Wannier-like Dualities in Quantum Lattice Models from Integrability}

\author{Rui-Dong Zhu}
\affiliation{Institute for Advanced Study \& School of Physical Science and Technology, Soochow University, Suzhou 215006, China}
\affiliation{Jiangsu Key Laboratory of Frontier Material Physics and Devices,\\ Soochow University, Suzhou 215006, China}

\begin{abstract}
The Kramers-Wannier duality introduces a well-known non-invertible symmetry in the critical transverse-field Ising model. In this work, we extend this concept to a broad class of quantum lattice models induced from integrability, providing explicit expressions for the Kramers-Wannier operators that can be systematically computed.

 \end{abstract}
\maketitle

\section{Introduction}

Non-invertible symmetries are playing more and more important roles in the classification of theories and phases in both high-energy and condensed matter physics. Its existence was first realized in 2d conformal field theories \cite{Verlinde:1988sn}, and then in field theories in arbitrary dimension, systematic approaches have been intensively developed recently through the category theory \cite{Fuchs:1993et,tensor-category} and symmetry topological field theory (symTFT) \cite{Witten:1998wy,Kaidi:2023maf} (see lecture notes \cite{Schafer-Nameki:2023jdn,Shao:2023gho}). In lattice models, the string net model \cite{Levin:2004mi,Hu:2015dga} serves the role of symTFT. However, unlike in the field theories that symmetries are represented by topological defects \cite{Frohlich:2004ef,Gaiotto:2014kfa,Chang:2018iay}, there still lacks a systematic way to find and construct non-invertible symmetry operators in quantum lattice models (c.f. fusion category approach \cite{Aasen:2020jwb} and quantum information aspect \cite{Okada:2024qmk} explored in the literature). 

One of the simplest examples is given by the celebrated Kramers-Wannier (KW) duality, originally found by Kramers and Wannier in the study of phase transition in the 2d classical Ising model \cite{Kramers:1941kn,Kramers:1941zz}. It can be mapped to a duality in the 1d quantum Ising model with a transverse magnetic field \cite{Aasen:2016dop}, usually referred to as the transverse-field Ising model (TFIM). The Hamiltonian is given by 
\begin{equation}
    {\cal H}_{\rm TFIM}=-J\sum_{i=1}^LX_iX_{i+1}-h\sum_{i=1}^LZ_i,\label{TFIM}
\end{equation}
and the KW duality exchanges the interaction term $\sum_i X_iX_{i+1}$ and $\sum_i Z_i$. It effectively switches the coupling constant $J$ and the magnetic field $h$, and relates the model in two opposite regions $J<h$ and $J>h$. A critical model arises at $J=h$, where the duality becomes a symmetry in the system. 
We denote the critical Hamiltonian with $J=h=1$ as ${\cal H}_{\rm cTFIM}$. A recent interesting idea proposed in \cite{Sinha:2025wqf} suggests a way to promote the invertible symmetries of an integrable model, whose Hamiltonian differs from ${\cal H}{\rm cTFIM}$ only by a boundary term, to the non-invertible symmetries present in ${\cal H}{\rm cTFIM}$. By modifying the boundary term, the standard symmetries in the integrable model become non-invertible, as each acquires a natural projector structure, in the critical TFIM. This offers a simple and elegant explanation for the emergence of non-invertible symmetry in the lattice TFIM. 

In this work, we extend this idea and show that in a large class of lattice models, non-invertible symmetries can be found systematically, with the Kramers-Wannier-like symmetry operators explicitly constructed. The article is organized as follows. In section \ref{s:review}, we first reproduce the result in \cite{Sinha:2025wqf} starting from a fermionic R-matrix. In section \ref{s:disguise}, we show that a large class of models equipped with the non-invertible symmetries can be built associated to a given fermionic R-matrix, and we present three explicit examples by restricting our focus on local Hamiltonians that can fit into the paradigm of the so-called free fermions in disguise. We then write down the explicit actions of Kramers-Wannier-like symmetries in these models in section \ref{s:KW-flow}, and we further argue that integrability can be partially preserved along a flow generated by these Kramers-Wannier operators. We conclude our paper by further providing examples built based on parafermions in section \ref{s:para}, and we propose a potential relation between the R-matrices and fusion rules (of Kramer-Wannier operators).  

\section{Kramers-Wannier Duality from Fermionic R-matrix}\label{s:review} 

We first prepare a set of an even number of Majorana fermions $\{\gamma_i\}_{i=1}^{2L}$ satisfying 
\begin{equation}
    \gamma_i=\gamma_i^\dagger,\quad \{\gamma_i,\gamma_j\}=2\delta_{i,j},
\end{equation}
and we further build an auxiliary space labeled by $0$ with the Majorana fermion in it denoted as $\gamma_0$. On each of the two Majorana-fermion spaces, there can be defined a fermionic R-matrix given by \cite{Sinha:2025wqf}
\begin{align}
    R_{jk}(u)=\frac{\gamma_j-\gamma_k}{\sqrt{2}}\left(1+i\tan(u)\gamma_j\gamma_k\right)\cr
    =\frac{1+i\tan(u)}{\sqrt{2}}\left(\gamma_j+f_0(u)\gamma_k\right),\label{r-maj}
\end{align}
where 
\begin{equation}
    f_0(u)=-\frac{1-i\tan(u)}{1+i\tan(u)}.
\end{equation}
It satisfies the standard Yang-Baxter equation, 
\begin{equation}
    R_{ij}(u-v)R_{ik}(u)R_{jk}(v)=R_{jk}(v)R_{ik}(u)R_{ij}(u-v).\label{YBE}
\end{equation}
It then follows straightforwardly from the standard algebraic Bethe ansatz approach \cite{Faddeev:1996iy} to use the transfer matrix constructed from the R-matrix, 
\begin{equation}
    \tau(u)={\rm tr}_0\prod_{j=1}^{2L}R_{0j}(u),
\end{equation}
to build an integrable lattice model. One can honestly take the trace on $0$-th space to compute $\tau$ by applying the Jordan-Wigner transformation including an auxiliary space, but one can also take a shortcut to simply extract out the even part of $\gamma_0$ in the expansion of $\prod_jR_{0j}$. From the commutativity of the transfer matrix, $[\tau(u),\tau(v)]=0$, an infinite set of conserved charges $\{\mathbb{Q}_i\}$ are obtained by identifying them as $\log\tau(u)=\sum_{i=0}^\infty\mathbb{Q}_iu^i$. Among them, two charge operators are in particular interesting for us, one is $\mathbb{Q}_0=\tau(0)$, and the other is the Hamiltonian of the system, usually chosen to be 
\begin{equation}
    {\cal H}:=\tau(0)^{-1}\left.\frac{{\rm d}\tau(u)}{{\rm d}u}\right|_{u=0}.
\end{equation}
We then obtain 
\begin{equation}
    \tau(0)=2\gamma_1(\gamma_1-\gamma_{2L})\dots (\gamma_1-\gamma_2),
\end{equation}
and
\begin{equation}
    {\cal H}=i\sum_{j=1}^{2L-1}\gamma_j\gamma_{j+1}-i\gamma_{2L}\gamma_1,
\end{equation}
with the normalization $\tau^\dagger(0)\tau(0)=2^{2L-1}$. One can then define a normalized charge operator ${\cal U}$, 
\begin{equation}
    {\cal U}:=2^{-L+1/2}\tau(0),\quad {\cal U}^{-1}=2^{-L+1/2}\tau^\dagger(0),
\end{equation}
and it acts on the Majorana fermions as a translation, 
\begin{equation}
    {\cal U}\gamma_j {\cal U}^{-1}=(-1)^{\delta_{j,2L}}\gamma_{j+1},\ {\rm with}\ \gamma_{2L+i}\equiv \gamma_i.
\end{equation}

To convert it to a spin chain model, we adopt the Jordan-Wigner transformation, 
\begin{align}
    \gamma_{2k-1}=\left[\prod_{j=1}^{k-1}Z_j\right]X_k,\quad \gamma_{2k}=\left[\prod_{j=1}^{k-1}Z_j\right]Y_k,
\end{align}
for $k=1,2,\dots,L$. The integrable Hamiltonian in the spin basis is found to be 
\begin{equation}
    {\cal H}=-\sum_{j=1}^LZ_j-\sum_{j=1}^{L-1}X_jX_{j+1}-{\cal P}X_LX_1,\label{ham-int}
\end{equation}
where ${\cal P}:=\prod_{j=1}^LZ_j$, and it satisfies $[{\cal H},{\cal P}]=0$ \footnote{This can alternatively be shown from the symmetry $\iota(\gamma_0)=-\gamma_0$ in the transfer matrix and $\{{\cal P},\gamma_j\}=0$ for $j=1,2,\dots,2L$.} and ${\cal P}^2=1$. It differs from the critical TFIM only by a boundary term, 
\begin{equation}
    {\cal H}-{\cal H}_{\rm cTFIM}=(1-{\cal P})X_LX_1.
\end{equation}
The operator $1-{\cal P}$ is a typical projector without the inverse. For any conserved charge $\mathbb{Q}_i$ in the integrbale system \eqref{ham-int}, one can define a non-invertible operator 
\begin{equation}
    {\cal Q}^+_i:=(1+{\cal P})\mathbb{Q}_i=\mathbb{Q}_i(1+{\cal P}),
\end{equation}
and it gives rise to a non-trivial non-invertible symmetry to the critical TFIM ${\cal H}_{\rm cTFIM}$, as one can see that ${\cal Q}^+_i$ commutes with ${\cal H}$ and kills the additional term $(1-{\cal P})X_LX_1$. In particular, for the shift operator ${\cal U}$, we have its action of ${\cal U}^+:=(1+{\cal P}){\cal U}={\cal U}(1+{\cal P})$ as 
\begin{equation}
    {\cal U}^+Z_j=(X_jX_{j+1}){\cal U}^+,\quad {\cal U}^+(X_jX_{j+1})=Z_{j+1}{\cal U}^+.
\end{equation}
This is exactly how the famous Kramers-Wannier duality acts in TFIM, and we call ${\cal U}^+$ the Kramers-Wannier operator. At the critical point $J=h$, it becomes a non-invertible symmetry of the system. In the original integrable models \eqref{ham-int} and the critical TFIM, we will refer to ${\cal U}$ and ${\cal U}^+$ as the Kramers-Wannier symmetry in this article. We note that this model has also been considered in \cite{Yan:2024yrw} as a deformation of the critical TFIM, which preserves the Kramers-Wannier symmetry. The higher charges ${\cal Q}^+_{i\geq 3}$ also give non-trivial invertible symmetries in the TFIM, but their actions are more involved and non-local, so we do not focus on them here. 

\section{Free Fermions in Disguise}\label{s:disguise}

In the algebraic Bethe ansatz approach, more integrable lattice models can be systematically generated by using the Lax matrices instead of the R-matrix.  
The Lax matrix 
\begin{equation}
    L_{0j}(u;a_j)\propto\gamma_0+a_jf_0(u)\gamma_j,\label{lax}
\end{equation}
for any number $a_j$ satisfies the so-called RLL relation with respect to the fermionic R-matrix \eqref{r-maj}, 
\begin{align}
    &R_{00'}(u-v)L_{0j}(u;a_j)L_{0'j}(v;a_j)\cr
    &=L_{0'j}(v;a_j)L_{0j}(u;a_j)R_{00'}(u-v).
\end{align}
This suggests that for any transfer matrix of the form, 
\begin{equation}
    \tau(u,\{a_i\}_{i=1}^{2L})={\rm tr}_0\prod_{j=1}^{2L}L_{0j}(u;a_j),\label{int-disguise}
\end{equation}
it also gives a huge class of integrable models that have a similar structure to the TFIM, but in general, they can have non-local (not nearest-neighbor) and complicated interactions. For simplicity, in this work, we only focus on a subset of Hamiltonians that are local and can be expressed in the form, 
\begin{equation}
    {\cal H}_{\rm disguise}=\sum_{i=1}^Mb_ih_i,\quad b_i\in\mathbb{C},
\end{equation}
where $h_i$'s contain even number of Majorana fermions and satisfy $[h_j,h_{k}]=0$ for $|j-k|>2$. Such systems are discovered in \cite{Fendley:2019sdx,Elman:2020eos} to be integrable and can be solved by rewriting the Hamiltonians in terms of free Dirac fermions. Note that we do not necessarily have $M=2L$, which fits in the case of critical TFIMs. When we further impose the condition that $\{h_j,h_{j+1}\}=0$, $[h_j,h_{k}]=0$ for $|j-k|>1$, the system becomes solvable in the sense of Jordan-Wigner. As we will see later, all the models constructed from the Lax matrix \eqref{lax} in this class involve only nearest-neighbor interactions when translated to the spin chain language (up to some boundary term containing ${\cal P}$). 

There are three classes of models found by solving the constraints $[h_j,h_{k}]=0$ ($|j-k|>2$). 
\begin{itemize}
    \item Model I: $a_i\neq 0$ for $i=1,2,\dots,2L$ and most $a_i$'s taking the value $a_i=\pm 1$. There are $2^{2L+1}$ such models. 
    \item Model II: A special model with $M=2L$ and only nearest-neighbor fermionic interaction $\gamma_j\gamma_{j+1}$ appears. It turns out that $a_i=\pm 1$ for $ i=1,2,\dots,2L$. There are in total $2^{2L}$ such models. 
    \item Model III: Allowing some of the coefficients $a_i=0$ knocks out $\gamma_i$ from the system. However, after the Jordan-Wigner transformation, it still involves all the sites of the corresponding spin chain. 
\end{itemize}
We note that we cannot have $a_j=\pm i$ for $^\forall j$, since $\tau(0)$ will be null in that case and does not have its inverse. 

As a concrete example, we provide three models respectively fitting in Model I, II, and III at $L=4$. Choosing $a_2=a_4=a_6=a_8=1$, $a_1=a_3=a_5=a_7=a$, we obtain the Hamiltonian of Model I up to some constant term given by \footnote{We only introduced one free parameter here for simplicity. In general, there can be arbitrarily many independent free parameters giving infinitely many different variations of models when $L\to\infty$. For example, at $L=4$, four parameters $a_{1,3,5,7}$ can be independent when choosing $a_{2,4,6,8}=1$ in Model I. }
\begin{align}
    {\cal H}_1=&i\frac{(1+a^2)^3}{8}\left(a\sum_{j=1}^{2L}:\gamma_j\gamma_{j+1}:-\frac{1-a^2}{2}\sum_{k=1}^{L}:\gamma_{2k}\gamma_{2k+2}:\right)\cr
    =&-\frac{a(1+a^2)^3}{8}\left[\sum_{i=1}^LZ_i+\sum_{i=1}^{L-1}X_iX_{i+1}\right.\cr
    &\left.-\frac{1-a^2}{2a}\sum_{i=1}^{L-1}X_iY_{i+1}+{\cal P}X_LX_1-\frac{1-a^2}{2a}{\cal P}X_LY_1\right].
\end{align}
When $a$ is close to $1$, it can be viewed as a perturbed critical TFIM, while if $a$ is close to $0$, it gives a model with only $XY$-type nearest-neighbor interaction. 
As an example of Model II, we choose $a_1=a_4=a_5=a_6=a_7=a_8=1$ and $a_2=a_3=-1$, then we have 
\begin{align}
    &{\cal H}_2=i\left(\sum_{j\notin \{1,3,\dots,L-1\}}:\gamma_j\gamma_{j+1}:-\sum_{k=1}^{L/2}:\gamma_{2k-1}\gamma_{2k}:\right)\cr
    &=-\left[-\sum_{i=1}^{L/2}Z_i+\sum_{i=L/2+1}^{L}Z_i+\sum_{i=2}^{L-1}X_iX_{i+1}\right]-{\cal P}X_LX_1.
\end{align}
In the above, we introduced a pseudo normal-ordering notation, $:\gamma_i\gamma_j:=:\gamma_j\gamma_i:=\gamma_i\gamma_j$ if $i<j$. We note that although we only show the parameter choices for $L=4$, but it is straightforward to generalize such models to any $L$ \footnote{For ${\cal H}_2$, we need even number of lattice sites to keep the exactly same form of the Hamiltonian, but the key idea is that we can freely change the signs of the coefficients before $X_iX_{i+1}$ or $Z_i$. }. For Model III, we randomly choose $a_1=a_3=a$, $a_2=a_4=a_7=1$, $a_5=a_8=-1$, and $a_6=0$ as an example. The resulting Hamiltonian is given by 
\begin{align}
    H_3=&\frac{i(1+a^2)}{8}\left[2a\gamma_1\gamma_8+2a\gamma_1\gamma_2+2a\gamma_2\gamma_3+2a\gamma_3\gamma_4\right.\cr
    &\left.+(1+a^2)\gamma_4\gamma_5+(1+a^2)\gamma_5\gamma_7-(1+a^2)\gamma_7\gamma_8\right]\cr
    =&-\frac{a(1+a^2)}{4}\left[Z_1+Z_2+X_1X_2+\frac{1+a^2}{2a}X_2X_3\right.\cr
    &\left.-\frac{1+a^2}{2a}Y_3X_4-\frac{1+a^2}{2a}Z_4+{\cal P}X_4X_1\right]
\end{align}
We observe that we always have ${\cal P}$ appearing in accompany with the boundary terms in the integrable models, and we will refer to the periodic model without ${\cal P}$ as the generalized critical TFIMs. The pattern of Model III is relatively complicated, and detailed analysis will only be presented in a longer version of this article.

\section{Kramers-Wannier-like Symmetry in Generalized Critical TFIMs and Dualities} \label{s:KW-flow}

Now we turn to the analysis of the Kramers-Wannier-like symmetry operator ${\cal U}^+$ in the generalized critical TFIMs. We normalize ${\cal U}_i\propto\tau(0)$ in ${\cal H}_{i=1,2}$ s.t. ${\cal U}^\dagger{\cal U}=1$. Their action differs very much from that of ${\cal U}$ in the original TFIM. 
\begin{align}
    &{\cal U}_1\gamma_{2k-1}{\cal U}^{-1}_1=\frac{2a}{1+a^2}\gamma_{2k}+\frac{1-a^2}{1+a^2}\gamma_{2k-1},\\
    &{\cal U}_1\gamma_{2k}{\cal U}^{-1}_1=(-1)^{\delta_{k,L}}\left(\frac{2a}{1+a^2}\gamma_{2k+1}-\frac{1-a^2}{1+a^2}\gamma_{2k+2}\right),\notag
\end{align}
\begin{equation}
    {\cal U}_2\gamma_{j}{\cal U}^{-1}_2=(-1)^{\delta_{j,1}+\delta_{j,3}+\delta_{j,2L}}\gamma_{j+1}.
\end{equation}
These operators are promoted to non-invertible symmetries again by multiplying $1+{\cal P}$ to them, 
\begin{equation}
    {\cal U}^+_i:={\cal U}_i(1+{\cal P})=(1+{\cal P}){\cal U}_i,
\end{equation}
and we will refer to them as the Kramers-Wannier operators. 
Then we have in the spin basis, 
\begin{align}
    &{\cal U}^+_1Z_j=\left[\frac{4a^2}{(1+a^2)^2}X_jX_{j+1}+\frac{(1-a^2)^2}{(1+a^2)^2}Y_jY_{j+1}\right.\cr
    &\quad \quad \left.-\frac{2a(1-a^2)}{(1+a^2)^2}(X_jY_{j+1}+Y_jX_{j+1})\right]{\cal U}^+_1,\cr
    &{\cal U}^+_1X_jX_{j+1}=Z_{j+1}{\cal U}^+_1,
\end{align}
\begin{align}
    &{\cal U}^+_1X_jY_{j+1}=\left[\frac{2a(1-a^2)}{(1+a^2)^2}Y_{j+1}Y_{j+2}\right.\cr
    &\quad -\frac{2a(1-a^2)}{(1+a^2)^2}X_{j+1}X_{j+2}-\frac{4a^2}{(1+a^2)^2}Y_{j+1}X_{j+2}\cr
    &\quad \left.+\frac{(1-a^2)^2}{(1+a^2)^2}X_{j+1}Y_{j+2}\right]{\cal U}^+_1.
\end{align}
One can easily check that the critical TFIM corresponding to ${\cal H}_1$ with ${\cal P}$ removed is invariant under the above action. A Kramers-Wannier-type duality can also be induced for the system, e.g. 
\begin{align}
    {\cal H}_{gTFIM_1}:=-J\sum_{i=1}^LX_iX_{i+1}-h\sum_{i=1}^LZ_i-gh\sum_{i=1}^LX_iY_{i+1},
\end{align}
and by solving $g=-\frac{1-a^2}{2a}$ for $a$, one finds a duality operator that exchanges $J\leftrightarrow h$ as in the original TFIM. 

Furthermore, one can create a flow with ${\cal U}^+_1$ for arbitrary $a$ to build a family of lattice models starting from a Hamiltonian of the generalized TFIM type (Similar idea has been explored in the literature \cite{Eck:2023gic,Cao:2025qnc} under the name of non-invertible mapping). With a pair of Hamiltonians $({\cal H},{\cal H}')$ satisfying ${\cal U}^+_1{\cal H}={\cal H}'{\cal U}^+_1$, eigenstates of ${\cal H}'$ can be obtained from those of ${\cal H}$ (denoted as $\ket{E}$) by acting ${\cal U}_1^+$ on them, 
\begin{equation}
    {\cal U}^+_1{\cal H}\ket{E}=E{\cal U}^+_1\ket{E}={\cal H}'{\cal U}^+_1\ket{E}.
\end{equation}
As long as ${\cal U}^+_1\ket{E}\neq 0$ (not annihilated by the projector), it gives an eigenstate with the same energy in ${\cal H}'$. 
For example, starting from ${\cal H}_{\rm TFIM}$, we obtain a series of systems sharing (at least part of) the energy spectrum with ${\cal H}_{\rm TFIM}$ by repeatedly acting ${\cal U}^+_1$ onto the Hamiltonian. Acting ${\cal U}^+_1$ once on ${\cal H}_{\rm TFIM}$, one finds the following Hamiltonian with a free parameter $a$, 
\begin{align}
    {\cal H}'_{\rm TFIM}(a)&=-J\sum_{i=1}^LZ_i-\frac{4a^2h}{(1+a^2)^2}\sum_{i=1}^LX_iX_{i+1}\cr
    &-\frac{(1-a^2)^2h}{(1+a^2)^2}\sum_{i=1}^LY_iY_{i+1}+\frac{2a(1-a^2)h}{(1+a^2)^2}\cr
    &\times \sum_{i=1}^L(X_iY_{i+1}+Y_iX_{i+1}).
\end{align}
In the critical model with $J=h$, its spectrum completely matches that of the TFIM \eqref{TFIM}, as no state is annihilated by the non-invertible symmetry ${\cal U}^+$, and therefore the projector $1+{\cal P}$. More generally, only the subspace specified by the projector $1+{\cal P}$ and the corresponding energy eigenvalues are shared by the models before and after the flow. Global symmetries, if not annihilated by ${\cal U}^+_1$, will also be inherited through the flow, so if the starting point is an integrable model, the integrability will be partially preserved in the subspace of $1+{\cal P}$. 

The action of ${\cal U}^+_2$ in the spin basis reads 
\begin{align}
    &{\cal U}^+_2Z_j=(-1)^{\theta(L/2-j)}X_jX_{j+1}{\cal U}^+_2,\cr
    &{\cal U}_2^+X_{j-1}X_{j}=(-1)^{\theta(L/2-j)}Z_{j}{\cal U}_2^+,
\end{align}
where $\theta(x)$ denotes the step function with $\theta(0):=1$. It implies that there is also a duality in the following model, 
\begin{align}
    {\cal H}_{gTFIM_2}:=-J\sum_{i=1}^LX_iX_{i+1}-h\sum_{i=1}^L(-1)^{\theta(L/2-i)}Z_i,
\end{align}
by exchanging $J\leftrightarrow h$. The flow created by ${\cal U}^+_2$ allows us to flip the signs of local interactions and magnetic fields while preserving part of the spectrum.

\section{Fusion Rules and Parafermions}\label{s:para}

So far, we have been considering critical models induced from the R-matrix of Majorana fermions \eqref{r-maj}. By construction, the Kramers-Wannier operator ${\cal U}_i^+$, when multiplied with its conjugate, becomes the projector $1+{\cal P}$. This can be summarized into the following fusion rule,
\begin{equation}
    ({\cal U}^\dagger_i)^+\times {\cal U}^+_i=1+{\cal P}.
\end{equation}

More generally, one can consider $\mathbb{Z}_p$ parafermions satisfying (for a root of unity element $\omega^p=1$, $\omega\neq 1$),
\begin{equation}
    \psi_i\psi_j=\omega \psi_j\psi_i,\quad i>j,\quad \psi_i^p=1,
\end{equation}
and try to repeat the whole procedure we did for Majorana fermions. The R-matrix can be found by Baxterizing the parafermion representation of the braid group shown in \cite{Cobanera:2013pua}. Let us write down the explicit form for the case $p=3$. 
\begin{equation}
    \check{R}_{jk}(u)=1+g_0(u)\left[\psi^\dagger_j\psi_k+\omega\psi_j\psi^\dagger_k\right],
\end{equation}
where we denoted $\psi^\dagger_j=\psi_j^2$,  
\begin{equation}
    g_0(u)=\frac{1}{2\omega}\left(\sqrt{3}\tan\left[\frac{\sqrt{3}\omega}{2}u+\frac{\pi}{6}\right]\right),
\end{equation}
and $\check{R}_{jk}$ is the R-matrix in the braided form satisfying the braided Yang-Baxter equation, 
\begin{equation}
    \check{R}_{ij}(u-v)\check{R}_{jk}(u)\check{R}_{ij}(v)=\check{R}_{jk}(v)\check{R}_{ij}(u)\check{R}_{jk}(u-v),
\end{equation}
which can be obtained from the original Yang-Baxter equation \eqref{YBE} by multiplying a permutation-like operator to $R_{jk}$. 
The local integrable Hamiltonian constructed from the R-matrix can be worked out as \footnote{One may also formally define a permutation operator $P^\omega_{12}:\ \psi_1\to \omega\psi_2,\psi_2\to \omega\psi_1$ to repeat the standard algebraic Bethe ansatz approach, but $P^\omega$ cannot be constructed only in terms of parafermions. See a longer version of this paper to appear for more details. }
\begin{equation}
    {\cal H}_{j(j+1)}\propto \left.\frac{{\rm d}\check{R}_{j(j+1)}}{{\rm d}u}\right|_{u=0}=\psi_j^\dagger\psi_{j+1}+h.c.
\end{equation}
We expect the similar form to hold for arbitrary (at least prime) $p$, and this is exactly the local interaction appearing in vector Potts/$p$-clock models. By applying a Jordan-Wigner-like transformation, 
\begin{equation}
    \psi_{2k-1}=\left(\prod_{j=1}^{k-1}U_j\right)V_k,\quad \psi_{2k}=\left(\prod_{j=1}^{k-1}U_j\right)V_kU_k,
\end{equation}
with 
\begin{equation}
    V=\left(\begin{matrix}
    0 & 1 & 0\\
    0 & 0 & 1\\
    1 & 0 & 0\\
    \end{matrix}\right),\quad U=\left(\begin{matrix}
    1 & 0 & 0\\
    0 & \omega & 0\\
    0 & 0 & \omega^2\\
    \end{matrix}\right),
\end{equation}
the integrable Hamiltonian is converted to 
\begin{align}
    {\cal H}_{\rm iVP}\propto\sum_{i=1}^LU_i+\sum_{i=1}^{L-1}V_iV_{i+1}^\dagger+{\cal O}V_LV^\dagger_1+h.c.\cr
    =:{\cal H}_{\rm cVP}+\left[(1-{\cal O})V_LV^\dagger_1+h.c.\right],
\end{align}
where ${\cal O}:=\prod_{i=1}^LU_i$. Similar R-matrix and integrable models for $\mathbb{Z}_3$ have been constructed in \cite{Yu-Ge} without careful analysis on the boundary condition. 

We again have some mismatch between the integrable model and the critical vector Potts model ${\cal H}_{\rm cVP}$ on the boundary. The extra terms contain either the operator $1-{\cal O}$ or its conjugate $1-{\cal O}^2$, and we note that both of them commute with the projector $1+{\cal O}+{\cal O}^2$. Therefore, for each conserved charge ${\mathbb Q}_i$ in the integrable model, we have a corresponding non-invertible symmetry given by the charge 
\begin{equation}
    {\cal Q}_i^{++}:=(1+{\cal O}+{\cal O}^2){\mathbb Q}_i={\mathbb Q}_i(1+{\cal O}+{\cal O}^2).
\end{equation}
We can then read off the fusion rule for the Kramer-Wannier-like duality in the $\mathbb{Z}_3$ vector Potts model as 
\begin{equation}
    ({\cal Q}^\dagger)^{++}\times {\cal Q}^{++}=1+{\cal O}+{\cal O}^2.
\end{equation}
The Kramers-Wannier symmetry operator has been constructed explicitly in \cite{Zhang:2020pco}. 
In general cases with arbitrary $p$, we further expect it to be modified to 
\begin{equation}
    ({\cal Q}^\dagger)^{++\dots +}\times {\cal Q}^{++\dots +}=1+\sum_{a=1}^{p-1}{\cal D}^a,\quad {\cal D}^p=1.
\end{equation}
Here we use the notation such that the signs $\sigma_a=\pm$ on the superscript stand for the projector $1+\sum_{a=1}^{p-1}\sigma_a{\cal D}^a$ multiplied to a given charge operator ${\cal Q}$. For $p=2$, ${\cal D}={\cal P}$, and for $p=3$, ${\cal D}={\cal O}$ in our previous examples. This agrees with the rule in the literature, e.g. \cite[(7.6)]{Aasen:2020jwb}. Replacing the R-matrix further with suitable Lax matrices is expected to give more abundant variants of critical Potts models, and some of them may coincide with those presented in \cite{Lahtinen:2017rjd} obtained from a CFT approach. 

One may also consider a system that consists of two decoupled 1d lattices. By using two R-matrices, $\check{R}^{(1)}$ and $\check{R}^{(2)}$, constructed from two completely independent sets of free (para)fermions, one can construct an R-matrix, 
\begin{equation}
    \check{R}_{jk}(u)=\check{R}_{jk}^{(1)}\check{R}_{jk}^{(2)},\label{R-dec}
\end{equation}
to build an integrable model with the Hamiltonian ${\cal H}={\cal H}^{(1)}+{\cal H}^{(2)}$. It commutes with three transfer matrices, respectively constructed from $\check{R}^{(1)}$, $\check{R}^{(2)}$ and $\check{R}$. Denoting the critical models associated to ${\cal H}$ and ${\cal H}^{(i)}$'s obtained by modifying the boundary term as ${\cal H}_c$ and ${\cal H}^{(i)}_c$, we see that three transfer matrices give rise to non-invertible symmetries that respectively act only on ${\cal H}^{(1)}_c$, ${\cal H}^{(2)}_c$ and simultaneously on ${\cal H}_c$. Although this is a trivial example, it enriches the fusion rules, and we expect that by further refining the R-matrix \eqref{R-dec} to include interactions between two decoupled systems (c.f. in the two-leg Heisenberg ladder model, the R-matrix is an SU(4) version \cite{ladder-XXX}), non-invertible symmetries in ladder models can also be explicitly constructed in a systematic way.

\section{Conclusion and Discussion}

In this article, we explored a simple and systematic way to construct 1d quantum lattice models with non-invertible symmetries by using Lax operators (with free parameters) associated to fermionic R-matrices. Kramers-Wannier operators with continuous parameters can be written down explicitly in these models, and they can alternatively be used to generate a flow (also known as a non-invertible mapping) starting from a solvable model to many other lattice models while (partially) preserving the integrability (i.e., global symmetries) in a subspace. Fusion rules for such non-invertible operators are proposed to be determined by the underlying fermionic R-matrix. The framework looks similar to that used in \cite{Eck:2023gic,Cao:2025qnc}, and it will be interesting to formulate a concrete connection in the future. 

Since the Kramers-Wannier operators act on the models obtained from the integrability by modifying the boundary term as non-invertible symmetries (rather than dualities), we expect all such models to be critical. Therefore, in the continuum limit, they will flow to some 2d CFTs. We plan to investigate them as future work and compare with related results in the literature \cite{Mong:2014ova,2015PhRvL.115w7203L,Lahtinen:2017rjd,Hung:2025gcp}. Deformations of critical models preserving the integrability \cite{8-vert1,8-vert2} or non-invertible symmetries \cite{2015PhRvB..92w5123R,Yan:2024yrw} may further reveal the hidden relation between these two subjects.

\subsection*{Acknowledgment} The author would like to thank Jin Chen, Yunfeng Jiang, Hao Zou, and Weiguang Cao for inspiring discussions. This work is partially supported by the National Natural Science Foundation of China (Grant No. 12105198).

\bibliography{bib}

\begin{thebibliography}{39}%
\makeatletter
\providecommand \@ifxundefined [1]{%
 \@ifx{#1\undefined}
}%
\providecommand \@ifnum [1]{%
 \ifnum #1\expandafter \@firstoftwo
 \else \expandafter \@secondoftwo
 \fi
}%
\providecommand \@ifx [1]{%
 \ifx #1\expandafter \@firstoftwo
 \else \expandafter \@secondoftwo
 \fi
}%
\providecommand \natexlab [1]{#1}%
\providecommand \enquote  [1]{``#1''}%
\providecommand \bibnamefont  [1]{#1}%
\providecommand \bibfnamefont [1]{#1}%
\providecommand \citenamefont [1]{#1}%
\providecommand \href@noop [0]{\@secondoftwo}%
\providecommand \href [0]{\begingroup \@sanitize@url \@href}%
\providecommand \@href[1]{\@@startlink{#1}\@@href}%
\providecommand \@@href[1]{\endgroup#1\@@endlink}%
\providecommand \@sanitize@url [0]{\catcode `\\12\catcode `\$12\catcode `\&12\catcode `\#12\catcode `\^12\catcode `\_12\catcode `\%12\relax}%
\providecommand \@@startlink[1]{}%
\providecommand \@@endlink[0]{}%
\providecommand \url  [0]{\begingroup\@sanitize@url \@url }%
\providecommand \@url [1]{\endgroup\@href {#1}{\urlprefix }}%
\providecommand \urlprefix  [0]{URL }%
\providecommand \Eprint [0]{\href }%
\providecommand \doibase [0]{http://dx.doi.org/}%
\providecommand \selectlanguage [0]{\@gobble}%
\providecommand \bibinfo  [0]{\@secondoftwo}%
\providecommand \bibfield  [0]{\@secondoftwo}%
\providecommand \translation [1]{[#1]}%
\providecommand \BibitemOpen [0]{}%
\providecommand \bibitemStop [0]{}%
\providecommand \bibitemNoStop [0]{.\EOS\space}%
\providecommand \EOS [0]{\spacefactor3000\relax}%
\providecommand \BibitemShut  [1]{\csname bibitem#1\endcsname}%
\let\auto@bib@innerbib\@empty
\bibitem [{\citenamefont {Verlinde}(1988)}]{Verlinde:1988sn}%
  \BibitemOpen
  \bibfield  {author} {\bibinfo {author} {\bibfnamefont {E.~P.}\ \bibnamefont {Verlinde}},\ }\href {\doibase 10.1016/0550-3213(88)90603-7} {\bibfield  {journal} {\bibinfo  {journal} {Nucl. Phys. B}\ }\textbf {\bibinfo {volume} {300}},\ \bibinfo {pages} {360} (\bibinfo {year} {1988})}\BibitemShut {NoStop}%
\bibitem [{\citenamefont {Fuchs}(1994)}]{Fuchs:1993et}%
  \BibitemOpen
  \bibfield  {author} {\bibinfo {author} {\bibfnamefont {J.}~\bibnamefont {Fuchs}},\ }\href {\doibase 10.1002/prop.2190420102} {\bibfield  {journal} {\bibinfo  {journal} {Fortsch. Phys.}\ }\textbf {\bibinfo {volume} {42}},\ \bibinfo {pages} {1} (\bibinfo {year} {1994})},\ \Eprint {http://arxiv.org/abs/hep-th/9306162}{arXiv:hep-th/9306162}\BibitemShut {NoStop}%
\bibitem [{\citenamefont {Etingof}\ \emph {et~al.}(2015)\citenamefont {Etingof}, \citenamefont {Gelaki}, \citenamefont {Nikshych},\ and\ \citenamefont {Ostrik}}]{tensor-category}%
  \BibitemOpen
  \bibfield  {author} {\bibinfo {author} {\bibfnamefont {P.}~\bibnamefont {Etingof}}, \bibinfo {author} {\bibfnamefont {S.}~\bibnamefont {Gelaki}}, \bibinfo {author} {\bibfnamefont {D.}~\bibnamefont {Nikshych}}, \ and\ \bibinfo {author} {\bibfnamefont {V.}~\bibnamefont {Ostrik}},\ }\href {\doibase 10.1090/surv/205} {\bibfield  {journal} {\bibinfo  {journal} {Mathematical Surveys and Monographs}\ }\textbf {\bibinfo {volume} {205}} (\bibinfo {year} {2015}),\ 10.1090/surv/205}\BibitemShut {NoStop}%
\bibitem [{\citenamefont {Witten}(1998)}]{Witten:1998wy}%
  \BibitemOpen
  \bibfield  {author} {\bibinfo {author} {\bibfnamefont {E.}~\bibnamefont {Witten}},\ }\href {\doibase 10.1088/1126-6708/1998/12/012} {\bibfield  {journal} {\bibinfo  {journal} {JHEP}\ }\textbf {\bibinfo {volume} {12}},\ \bibinfo {pages} {012} (\bibinfo {year} {1998})},\ \Eprint {http://arxiv.org/abs/hep-th/9812012}{arXiv:hep-th/9812012}\BibitemShut {NoStop}%
\bibitem [{\citenamefont {Kaidi}\ \emph {et~al.}(2023)\citenamefont {Kaidi}, \citenamefont {Nardoni}, \citenamefont {Zafrir},\ and\ \citenamefont {Zheng}}]{Kaidi:2023maf}%
  \BibitemOpen
  \bibfield  {author} {\bibinfo {author} {\bibfnamefont {J.}~\bibnamefont {Kaidi}}, \bibinfo {author} {\bibfnamefont {E.}~\bibnamefont {Nardoni}}, \bibinfo {author} {\bibfnamefont {G.}~\bibnamefont {Zafrir}}, \ and\ \bibinfo {author} {\bibfnamefont {Y.}~\bibnamefont {Zheng}},\ }\href {\doibase 10.1007/JHEP10(2023)053} {\bibfield  {journal} {\bibinfo  {journal} {JHEP}\ }\textbf {\bibinfo {volume} {10}},\ \bibinfo {pages} {053} (\bibinfo {year} {2023})},\ \Eprint {http://arxiv.org/abs/2301.07112}{arXiv:2301.07112 [hep-th]}\BibitemShut {NoStop}%
\bibitem [{\citenamefont {Schafer-Nameki}(2024)}]{Schafer-Nameki:2023jdn}%
  \BibitemOpen
  \bibfield  {author} {\bibinfo {author} {\bibfnamefont {S.}~\bibnamefont {Schafer-Nameki}},\ }\href {\doibase 10.1016/j.physrep.2024.01.007} {\bibfield  {journal} {\bibinfo  {journal} {Phys. Rept.}\ }\textbf {\bibinfo {volume} {1063}},\ \bibinfo {pages} {1} (\bibinfo {year} {2024})},\ \Eprint {http://arxiv.org/abs/2305.18296}{arXiv:2305.18296 [hep-th]}\BibitemShut {NoStop}%
\bibitem [{\citenamefont {Shao}(2023)}]{Shao:2023gho}%
  \BibitemOpen
  \bibfield  {author} {\bibinfo {author} {\bibfnamefont {S.-H.}\ \bibnamefont {Shao}},\ }\href@noop {} {\  (\bibinfo {year} {2023})},\ \Eprint {http://arxiv.org/abs/2308.00747}{arXiv:2308.00747 [hep-th]}\BibitemShut {NoStop}%
\bibitem [{\citenamefont {Levin}\ and\ \citenamefont {Wen}(2005)}]{Levin:2004mi}%
  \BibitemOpen
  \bibfield  {author} {\bibinfo {author} {\bibfnamefont {M.~A.}\ \bibnamefont {Levin}}\ and\ \bibinfo {author} {\bibfnamefont {X.-G.}\ \bibnamefont {Wen}},\ }\href {\doibase 10.1103/PhysRevB.71.045110} {\bibfield  {journal} {\bibinfo  {journal} {Phys. Rev. B}\ }\textbf {\bibinfo {volume} {71}},\ \bibinfo {pages} {045110} (\bibinfo {year} {2005})},\ \Eprint {http://arxiv.org/abs/cond-mat/0404617}{arXiv:cond-mat/0404617}\BibitemShut {NoStop}%
\bibitem [{\citenamefont {Hu}\ \emph {et~al.}(2018)\citenamefont {Hu}, \citenamefont {Geer},\ and\ \citenamefont {Wu}}]{Hu:2015dga}%
  \BibitemOpen
  \bibfield  {author} {\bibinfo {author} {\bibfnamefont {Y.}~\bibnamefont {Hu}}, \bibinfo {author} {\bibfnamefont {N.}~\bibnamefont {Geer}}, \ and\ \bibinfo {author} {\bibfnamefont {Y.-S.}\ \bibnamefont {Wu}},\ }\href {\doibase 10.1103/PhysRevB.97.195154} {\bibfield  {journal} {\bibinfo  {journal} {Phys. Rev. B}\ }\textbf {\bibinfo {volume} {97}},\ \bibinfo {pages} {195154} (\bibinfo {year} {2018})},\ \Eprint {http://arxiv.org/abs/1502.03433}{arXiv:1502.03433 [cond-mat.str-el]}\BibitemShut {NoStop}%
\bibitem [{\citenamefont {Frohlich}\ \emph {et~al.}(2004)\citenamefont {Frohlich}, \citenamefont {Fuchs}, \citenamefont {Runkel},\ and\ \citenamefont {Schweigert}}]{Frohlich:2004ef}%
  \BibitemOpen
  \bibfield  {author} {\bibinfo {author} {\bibfnamefont {J.}~\bibnamefont {Frohlich}}, \bibinfo {author} {\bibfnamefont {J.}~\bibnamefont {Fuchs}}, \bibinfo {author} {\bibfnamefont {I.}~\bibnamefont {Runkel}}, \ and\ \bibinfo {author} {\bibfnamefont {C.}~\bibnamefont {Schweigert}},\ }\href {\doibase 10.1103/PhysRevLett.93.070601} {\bibfield  {journal} {\bibinfo  {journal} {Phys. Rev. Lett.}\ }\textbf {\bibinfo {volume} {93}},\ \bibinfo {pages} {070601} (\bibinfo {year} {2004})},\ \Eprint {http://arxiv.org/abs/cond-mat/0404051}{arXiv:cond-mat/0404051}\BibitemShut {NoStop}%
\bibitem [{\citenamefont {Gaiotto}\ \emph {et~al.}(2015)\citenamefont {Gaiotto}, \citenamefont {Kapustin}, \citenamefont {Seiberg},\ and\ \citenamefont {Willett}}]{Gaiotto:2014kfa}%
  \BibitemOpen
  \bibfield  {author} {\bibinfo {author} {\bibfnamefont {D.}~\bibnamefont {Gaiotto}}, \bibinfo {author} {\bibfnamefont {A.}~\bibnamefont {Kapustin}}, \bibinfo {author} {\bibfnamefont {N.}~\bibnamefont {Seiberg}}, \ and\ \bibinfo {author} {\bibfnamefont {B.}~\bibnamefont {Willett}},\ }\href {\doibase 10.1007/JHEP02(2015)172} {\bibfield  {journal} {\bibinfo  {journal} {JHEP}\ }\textbf {\bibinfo {volume} {02}},\ \bibinfo {pages} {172} (\bibinfo {year} {2015})},\ \Eprint {http://arxiv.org/abs/1412.5148}{arXiv:1412.5148 [hep-th]}\BibitemShut {NoStop}%
\bibitem [{\citenamefont {Chang}\ \emph {et~al.}(2019)\citenamefont {Chang}, \citenamefont {Lin}, \citenamefont {Shao}, \citenamefont {Wang},\ and\ \citenamefont {Yin}}]{Chang:2018iay}%
  \BibitemOpen
  \bibfield  {author} {\bibinfo {author} {\bibfnamefont {C.-M.}\ \bibnamefont {Chang}}, \bibinfo {author} {\bibfnamefont {Y.-H.}\ \bibnamefont {Lin}}, \bibinfo {author} {\bibfnamefont {S.-H.}\ \bibnamefont {Shao}}, \bibinfo {author} {\bibfnamefont {Y.}~\bibnamefont {Wang}}, \ and\ \bibinfo {author} {\bibfnamefont {X.}~\bibnamefont {Yin}},\ }\href {\doibase 10.1007/JHEP01(2019)026} {\bibfield  {journal} {\bibinfo  {journal} {JHEP}\ }\textbf {\bibinfo {volume} {01}},\ \bibinfo {pages} {026} (\bibinfo {year} {2019})},\ \Eprint {http://arxiv.org/abs/1802.04445}{arXiv:1802.04445 [hep-th]}\BibitemShut {NoStop}%
\bibitem [{\citenamefont {Aasen}\ \emph {et~al.}(2020)\citenamefont {Aasen}, \citenamefont {Fendley},\ and\ \citenamefont {Mong}}]{Aasen:2020jwb}%
  \BibitemOpen
  \bibfield  {author} {\bibinfo {author} {\bibfnamefont {D.}~\bibnamefont {Aasen}}, \bibinfo {author} {\bibfnamefont {P.}~\bibnamefont {Fendley}}, \ and\ \bibinfo {author} {\bibfnamefont {R.~S.~K.}\ \bibnamefont {Mong}},\ }\href@noop {} {\  (\bibinfo {year} {2020})},\ \Eprint {http://arxiv.org/abs/2008.08598}{arXiv:2008.08598 [cond-mat.stat-mech]}\BibitemShut {NoStop}%
\bibitem [{\citenamefont {Okada}\ and\ \citenamefont {Tachikawa}(2024)}]{Okada:2024qmk}%
  \BibitemOpen
  \bibfield  {author} {\bibinfo {author} {\bibfnamefont {M.}~\bibnamefont {Okada}}\ and\ \bibinfo {author} {\bibfnamefont {Y.}~\bibnamefont {Tachikawa}},\ }\href {\doibase 10.1103/PhysRevLett.133.191602} {\bibfield  {journal} {\bibinfo  {journal} {Phys. Rev. Lett.}\ }\textbf {\bibinfo {volume} {133}},\ \bibinfo {pages} {191602} (\bibinfo {year} {2024})},\ \Eprint {http://arxiv.org/abs/2403.20062}{arXiv:2403.20062 [hep-th]}\BibitemShut {NoStop}%
\bibitem [{\citenamefont {Kramers}\ and\ \citenamefont {Wannier}(1941{\natexlab{a}})}]{Kramers:1941kn}%
  \BibitemOpen
  \bibfield  {author} {\bibinfo {author} {\bibfnamefont {H.~A.}\ \bibnamefont {Kramers}}\ and\ \bibinfo {author} {\bibfnamefont {G.~H.}\ \bibnamefont {Wannier}},\ }\href {\doibase 10.1103/PhysRev.60.252} {\bibfield  {journal} {\bibinfo  {journal} {Phys. Rev.}\ }\textbf {\bibinfo {volume} {60}},\ \bibinfo {pages} {252} (\bibinfo {year} {1941}{\natexlab{a}})}\BibitemShut {NoStop}%
\bibitem [{\citenamefont {Kramers}\ and\ \citenamefont {Wannier}(1941{\natexlab{b}})}]{Kramers:1941zz}%
  \BibitemOpen
  \bibfield  {author} {\bibinfo {author} {\bibfnamefont {H.~A.}\ \bibnamefont {Kramers}}\ and\ \bibinfo {author} {\bibfnamefont {G.~H.}\ \bibnamefont {Wannier}},\ }\href {\doibase 10.1103/PhysRev.60.263} {\bibfield  {journal} {\bibinfo  {journal} {Phys. Rev.}\ }\textbf {\bibinfo {volume} {60}},\ \bibinfo {pages} {263} (\bibinfo {year} {1941}{\natexlab{b}})}\BibitemShut {NoStop}%
\bibitem [{\citenamefont {Aasen}\ \emph {et~al.}(2016)\citenamefont {Aasen}, \citenamefont {Mong},\ and\ \citenamefont {Fendley}}]{Aasen:2016dop}%
  \BibitemOpen
  \bibfield  {author} {\bibinfo {author} {\bibfnamefont {D.}~\bibnamefont {Aasen}}, \bibinfo {author} {\bibfnamefont {R.~S.~K.}\ \bibnamefont {Mong}}, \ and\ \bibinfo {author} {\bibfnamefont {P.}~\bibnamefont {Fendley}},\ }\href {\doibase 10.1088/1751-8113/49/35/354001} {\bibfield  {journal} {\bibinfo  {journal} {J. Phys. A}\ }\textbf {\bibinfo {volume} {49}},\ \bibinfo {pages} {354001} (\bibinfo {year} {2016})},\ \Eprint {http://arxiv.org/abs/1601.07185}{arXiv:1601.07185 [cond-mat.stat-mech]}\BibitemShut {NoStop}%
\bibitem [{\citenamefont {Sinha}\ \emph {et~al.}(2025)\citenamefont {Sinha}, \citenamefont {Justin}, \citenamefont {Padmanabhan},\ and\ \citenamefont {Korepin}}]{Sinha:2025wqf}%
  \BibitemOpen
  \bibfield  {author} {\bibinfo {author} {\bibfnamefont {A.}~\bibnamefont {Sinha}}, \bibinfo {author} {\bibfnamefont {T.}~\bibnamefont {Justin}}, \bibinfo {author} {\bibfnamefont {P.}~\bibnamefont {Padmanabhan}}, \ and\ \bibinfo {author} {\bibfnamefont {V.}~\bibnamefont {Korepin}},\ }\href@noop {} {\  (\bibinfo {year} {2025})},\ \Eprint {http://arxiv.org/abs/2506.03668}{arXiv:2506.03668 [hep-th]}\BibitemShut {NoStop}%
\bibitem [{\citenamefont {Faddeev}(1996)}]{Faddeev:1996iy}%
  \BibitemOpen
  \bibfield  {author} {\bibinfo {author} {\bibfnamefont {L.~D.}\ \bibnamefont {Faddeev}},\ }in\ \href@noop {} {\emph {\bibinfo {booktitle} {{Les Houches School of Physics: Astrophysical Sources of Gravitational Radiation}}}}\ (\bibinfo {year} {1996})\ pp.\ \bibinfo {pages} {149--219},\ \Eprint {http://arxiv.org/abs/hep-th/9605187}{arXiv:hep-th/9605187}\BibitemShut {NoStop}%
\bibitem [{Note1()}]{Note1}%
  \BibitemOpen
  \bibinfo {note} {This can alternatively be shown from the symmetry $\iota (\gamma _0)=-\gamma _0$ in the transfer matrix and $\{{\protect \cal P},\gamma _j\}=0$ for $j=1,2,\protect \dots ,2L$.}\BibitemShut {Stop}%
\bibitem [{\citenamefont {Yan}\ \emph {et~al.}(2024)\citenamefont {Yan}, \citenamefont {Konik},\ and\ \citenamefont {Mitra}}]{Yan:2024yrw}%
  \BibitemOpen
  \bibfield  {author} {\bibinfo {author} {\bibfnamefont {F.}~\bibnamefont {Yan}}, \bibinfo {author} {\bibfnamefont {R.}~\bibnamefont {Konik}}, \ and\ \bibinfo {author} {\bibfnamefont {A.}~\bibnamefont {Mitra}},\ }\href@noop {} {\  (\bibinfo {year} {2024})},\ \Eprint {http://arxiv.org/abs/2410.17317}{arXiv:2410.17317 [cond-mat.str-el]}\BibitemShut {NoStop}%
\bibitem [{\citenamefont {Fendley}(2019)}]{Fendley:2019sdx}%
  \BibitemOpen
  \bibfield  {author} {\bibinfo {author} {\bibfnamefont {P.}~\bibnamefont {Fendley}},\ }\href {\doibase 10.1088/1751-8121/ab305d} {\bibfield  {journal} {\bibinfo  {journal} {J. Phys. A}\ }\textbf {\bibinfo {volume} {52}},\ \bibinfo {pages} {335002} (\bibinfo {year} {2019})},\ \Eprint {http://arxiv.org/abs/1901.08078}{arXiv:1901.08078 [cond-mat.stat-mech]}\BibitemShut {NoStop}%
\bibitem [{\citenamefont {Elman}\ \emph {et~al.}(2021)\citenamefont {Elman}, \citenamefont {Chapman},\ and\ \citenamefont {Flammia}}]{Elman:2020eos}%
  \BibitemOpen
  \bibfield  {author} {\bibinfo {author} {\bibfnamefont {S.~J.}\ \bibnamefont {Elman}}, \bibinfo {author} {\bibfnamefont {A.}~\bibnamefont {Chapman}}, \ and\ \bibinfo {author} {\bibfnamefont {S.~T.}\ \bibnamefont {Flammia}},\ }\href {\doibase 10.1007/s00220-021-04220-w} {\bibfield  {journal} {\bibinfo  {journal} {Commun. Math. Phys.}\ }\textbf {\bibinfo {volume} {388}},\ \bibinfo {pages} {969} (\bibinfo {year} {2021})},\ \Eprint {http://arxiv.org/abs/2012.07857}{arXiv:2012.07857 [quant-ph]}\BibitemShut {NoStop}%
\bibitem [{Note2()}]{Note2}%
  \BibitemOpen
  \bibinfo {note} {We only introduced one free parameter here for simplicity. In general, there can be arbitrarily many independent free parameters giving infinitely many different variations of models when $L\to \infty $. For example, at $L=4$, four parameters $a_{1,3,5,7}$ can be independent when choosing $a_{2,4,6,8}=1$ in Model I.}\BibitemShut {Stop}%
\bibitem [{Note3()}]{Note3}%
  \BibitemOpen
  \bibinfo {note} {For ${\protect \cal H}_2$, we need even number of lattice sites to keep the exactly same form of the Hamiltonian, but the key idea is that we can freely change the signs of the coefficients before $X_iX_{i+1}$ or $Z_i$.}\BibitemShut {Stop}%
\bibitem [{\citenamefont {Eck}\ and\ \citenamefont {Fendley}(2024)}]{Eck:2023gic}%
  \BibitemOpen
  \bibfield  {author} {\bibinfo {author} {\bibfnamefont {L.}~\bibnamefont {Eck}}\ and\ \bibinfo {author} {\bibfnamefont {P.}~\bibnamefont {Fendley}},\ }\href {\doibase 10.21468/SciPostPhys.16.5.127} {\bibfield  {journal} {\bibinfo  {journal} {SciPost Phys.}\ }\textbf {\bibinfo {volume} {16}},\ \bibinfo {pages} {127} (\bibinfo {year} {2024})},\ \Eprint {http://arxiv.org/abs/2302.14081}{arXiv:2302.14081 [cond-mat.stat-mech]}\BibitemShut {NoStop}%
\bibitem [{\citenamefont {Cao}\ \emph {et~al.}(2025)\citenamefont {Cao}, \citenamefont {Miao},\ and\ \citenamefont {Yamazaki}}]{Cao:2025qnc}%
  \BibitemOpen
  \bibfield  {author} {\bibinfo {author} {\bibfnamefont {W.}~\bibnamefont {Cao}}, \bibinfo {author} {\bibfnamefont {Y.}~\bibnamefont {Miao}}, \ and\ \bibinfo {author} {\bibfnamefont {M.}~\bibnamefont {Yamazaki}},\ }\href@noop {} {\  (\bibinfo {year} {2025})},\ \Eprint {http://arxiv.org/abs/2501.12514}{arXiv:2501.12514 [cond-mat.str-el]}\BibitemShut {NoStop}%
\bibitem [{\citenamefont {Cobanera}\ and\ \citenamefont {Ortiz}(2014)}]{Cobanera:2013pua}%
  \BibitemOpen
  \bibfield  {author} {\bibinfo {author} {\bibfnamefont {E.}~\bibnamefont {Cobanera}}\ and\ \bibinfo {author} {\bibfnamefont {G.}~\bibnamefont {Ortiz}},\ }\href {\doibase 10.1103/PhysRevA.89.012328} {\bibfield  {journal} {\bibinfo  {journal} {Phys. Rev. A}\ }\textbf {\bibinfo {volume} {89}},\ \bibinfo {pages} {012328} (\bibinfo {year} {2014})},\ \bibinfo {note} {[Erratum: Phys.Rev.A 91, 059901 (2015)]},\ \Eprint {http://arxiv.org/abs/1307.6214}{arXiv:1307.6214 [cond-mat.stat-mech]}\BibitemShut {NoStop}%
\bibitem [{Note4()}]{Note4}%
  \BibitemOpen
  \bibinfo {note} {One may also formally define a permutation operator $P^\omega _{12}:\ \psi _1\to \omega \psi _2,\psi _2\to \omega \psi _1$ to repeat the standard algebraic Bethe ansatz approach, but $P^\omega $ cannot be constructed only in terms of parafermions. See a longer version of this paper to appear for more details.}\BibitemShut {Stop}%
\bibitem [{\citenamefont {{Yu}}\ and\ \citenamefont {{Ge}}(2016)}]{Yu-Ge}%
  \BibitemOpen
  \bibfield  {author} {\bibinfo {author} {\bibfnamefont {L.-W.}\ \bibnamefont {{Yu}}}\ and\ \bibinfo {author} {\bibfnamefont {M.-L.}\ \bibnamefont {{Ge}}},\ }\href {\doibase 10.1038/srep21497} {\bibfield  {journal} {\bibinfo  {journal} {Scientific Reports}\ }\textbf {\bibinfo {volume} {6}},\ \bibinfo {eid} {21497} (\bibinfo {year} {2016})},\ \Eprint {http://arxiv.org/abs/1507.05269}{arXiv:1507.05269 [quant-ph]}\BibitemShut {NoStop}%
\bibitem [{\citenamefont {Zhang}\ and\ \citenamefont {Sierra}(2020)}]{Zhang:2020pco}%
  \BibitemOpen
  \bibfield  {author} {\bibinfo {author} {\bibfnamefont {H.-C.}\ \bibnamefont {Zhang}}\ and\ \bibinfo {author} {\bibfnamefont {G.}~\bibnamefont {Sierra}},\ }\href {\doibase 10.1007/JHEP05(2025)157} {\bibfield  {journal} {\bibinfo  {journal} {JHEP}\ }\textbf {\bibinfo {volume} {25}},\ \bibinfo {pages} {157} (\bibinfo {year} {2020})},\ \Eprint {http://arxiv.org/abs/2410.06727}{arXiv:2410.06727 [cond-mat.stat-mech]}\BibitemShut {NoStop}%
\bibitem [{\citenamefont {Lahtinen}\ \emph {et~al.}(2021)\citenamefont {Lahtinen}, \citenamefont {Mansson},\ and\ \citenamefont {Ardonne}}]{Lahtinen:2017rjd}%
  \BibitemOpen
  \bibfield  {author} {\bibinfo {author} {\bibfnamefont {V.}~\bibnamefont {Lahtinen}}, \bibinfo {author} {\bibfnamefont {T.}~\bibnamefont {Mansson}}, \ and\ \bibinfo {author} {\bibfnamefont {E.}~\bibnamefont {Ardonne}},\ }\href {\doibase 10.21468/SciPostPhysCore.4.2.014} {\bibfield  {journal} {\bibinfo  {journal} {SciPost Phys. Core}\ }\textbf {\bibinfo {volume} {4}},\ \bibinfo {pages} {014} (\bibinfo {year} {2021})},\ \Eprint {http://arxiv.org/abs/1709.04259}{arXiv:1709.04259 [cond-mat.str-el]}\BibitemShut {NoStop}%
\bibitem [{\citenamefont {{Batchelor}}\ \emph {et~al.}(2007)\citenamefont {{Batchelor}}, \citenamefont {{Guan}}, \citenamefont {{Oelkers}},\ and\ \citenamefont {{Tsuboi}}}]{ladder-XXX}%
  \BibitemOpen
  \bibfield  {author} {\bibinfo {author} {\bibfnamefont {M.~T.}\ \bibnamefont {{Batchelor}}}, \bibinfo {author} {\bibfnamefont {X.~W.}\ \bibnamefont {{Guan}}}, \bibinfo {author} {\bibfnamefont {N.}~\bibnamefont {{Oelkers}}}, \ and\ \bibinfo {author} {\bibfnamefont {Z.}~\bibnamefont {{Tsuboi}}},\ }\href {\doibase 10.1080/00018730701265383} {\bibfield  {journal} {\bibinfo  {journal} {Advances in Physics}\ }\textbf {\bibinfo {volume} {56}},\ \bibinfo {pages} {465} (\bibinfo {year} {2007})},\ \Eprint {http://arxiv.org/abs/cond-mat/0512489}{arXiv:cond-mat/0512489 [cond-mat.stat-mech]}\BibitemShut {NoStop}%
\bibitem [{\citenamefont {Mong}\ \emph {et~al.}(2014)\citenamefont {Mong}, \citenamefont {Clarke}, \citenamefont {Alicea}, \citenamefont {Lindner},\ and\ \citenamefont {Fendley}}]{Mong:2014ova}%
  \BibitemOpen
  \bibfield  {author} {\bibinfo {author} {\bibfnamefont {R.~S.~K.}\ \bibnamefont {Mong}}, \bibinfo {author} {\bibfnamefont {D.~J.}\ \bibnamefont {Clarke}}, \bibinfo {author} {\bibfnamefont {J.}~\bibnamefont {Alicea}}, \bibinfo {author} {\bibfnamefont {N.~H.}\ \bibnamefont {Lindner}}, \ and\ \bibinfo {author} {\bibfnamefont {P.}~\bibnamefont {Fendley}},\ }\href {\doibase 10.1088/1751-8113/47/45/452001} {\bibfield  {journal} {\bibinfo  {journal} {J. Phys. A}\ }\textbf {\bibinfo {volume} {47}},\ \bibinfo {pages} {452001} (\bibinfo {year} {2014})},\ \Eprint {http://arxiv.org/abs/1406.0846}{arXiv:1406.0846 [cond-mat.stat-mech]}\BibitemShut {NoStop}%
\bibitem [{\citenamefont {{Lahtinen}}\ and\ \citenamefont {{Ardonne}}(2015)}]{2015PhRvL.115w7203L}%
  \BibitemOpen
  \bibfield  {author} {\bibinfo {author} {\bibfnamefont {V.}~\bibnamefont {{Lahtinen}}}\ and\ \bibinfo {author} {\bibfnamefont {E.}~\bibnamefont {{Ardonne}}},\ }\href {\doibase 10.1103/PhysRevLett.115.237203} {\bibfield  {journal} {\bibinfo  {journal} {\prl}\ }\textbf {\bibinfo {volume} {115}},\ \bibinfo {eid} {237203} (\bibinfo {year} {2015})},\ \Eprint {http://arxiv.org/abs/1504.07044}{arXiv:1504.07044 [cond-mat.str-el]}\BibitemShut {NoStop}%
\bibitem [{\citenamefont {Hung}\ \emph {et~al.}(2025)\citenamefont {Hung}, \citenamefont {Ji}, \citenamefont {Shen}, \citenamefont {Wan},\ and\ \citenamefont {Zhao}}]{Hung:2025gcp}%
  \BibitemOpen
  \bibfield  {author} {\bibinfo {author} {\bibfnamefont {L.-Y.}\ \bibnamefont {Hung}}, \bibinfo {author} {\bibfnamefont {K.}~\bibnamefont {Ji}}, \bibinfo {author} {\bibfnamefont {C.}~\bibnamefont {Shen}}, \bibinfo {author} {\bibfnamefont {Y.}~\bibnamefont {Wan}}, \ and\ \bibinfo {author} {\bibfnamefont {Y.}~\bibnamefont {Zhao}},\ }\href@noop {} {\  (\bibinfo {year} {2025})},\ \Eprint {http://arxiv.org/abs/2506.05324}{arXiv:2506.05324 [cond-mat.str-el]}\BibitemShut {NoStop}%
\bibitem [{\citenamefont {Wu}(1971)}]{8-vert1}%
  \BibitemOpen
  \bibfield  {author} {\bibinfo {author} {\bibfnamefont {F.~W.}\ \bibnamefont {Wu}},\ }\href {\doibase 10.1103/PhysRevB.4.2312} {\bibfield  {journal} {\bibinfo  {journal} {Phys. Rev. B}\ }\textbf {\bibinfo {volume} {4}},\ \bibinfo {pages} {2312} (\bibinfo {year} {1971})}\BibitemShut {NoStop}%
\bibitem [{\citenamefont {Kadanoff}\ and\ \citenamefont {Wegner}(1971)}]{8-vert2}%
  \BibitemOpen
  \bibfield  {author} {\bibinfo {author} {\bibfnamefont {L.~P.}\ \bibnamefont {Kadanoff}}\ and\ \bibinfo {author} {\bibfnamefont {F.~J.}\ \bibnamefont {Wegner}},\ }\href {\doibase 10.1103/PhysRevB.4.3989} {\bibfield  {journal} {\bibinfo  {journal} {Phys. Rev. B}\ }\textbf {\bibinfo {volume} {4}},\ \bibinfo {pages} {3989} (\bibinfo {year} {1971})}\BibitemShut {NoStop}%
\bibitem [{\citenamefont {{Rahmani}}\ \emph {et~al.}(2015)\citenamefont {{Rahmani}}, \citenamefont {{Zhu}}, \citenamefont {{Franz}},\ and\ \citenamefont {{Affleck}}}]{2015PhRvB..92w5123R}%
  \BibitemOpen
  \bibfield  {author} {\bibinfo {author} {\bibfnamefont {A.}~\bibnamefont {{Rahmani}}}, \bibinfo {author} {\bibfnamefont {X.}~\bibnamefont {{Zhu}}}, \bibinfo {author} {\bibfnamefont {M.}~\bibnamefont {{Franz}}}, \ and\ \bibinfo {author} {\bibfnamefont {I.}~\bibnamefont {{Affleck}}},\ }\href {\doibase 10.1103/PhysRevB.92.235123} {\bibfield  {journal} {\bibinfo  {journal} {\prb}\ }\textbf {\bibinfo {volume} {92}},\ \bibinfo {eid} {235123} (\bibinfo {year} {2015})},\ \Eprint {http://arxiv.org/abs/1505.03966}{arXiv:1505.03966 [cond-mat.str-el]}\BibitemShut {NoStop}%
\end{thebibliography}%
\bibliographystyle{apsrev4-2}

\end{document}